\documentclass[debug,overfull]{epl}
\usepackage{graphicx}
\usepackage{dcolumn}
\usepackage{bm}
\usepackage{epsfig}
\usepackage{color}
\usepackage{subeqnarray}
\usepackage{amsmath}
\definecolor{grey}{gray}{0.75}

\title{Optical response to magnetic ordering in \chem{Pr Mn O_3}: the relevance of the double exchange interaction}
\shorttitle{Optical response to magnetic ordering in \chem{Pr Mn O_3}}
\author{R. Sopracase \and G. Gruener \and C. Autret - Lambert \and V. Ta Phuoc \and V. Briz\'e \and J. C. Soret}
\institute{LEMA UMR 6157 CNRS-CEA, Universit\'{e} Fran\c{c}ois Rabelais, Parc de Grandmont, 37200 Tours, France}
 {Manganites}
 {Optical properties}

\begin{document}

\maketitle

\begin{abstract}
The optical conductivity of the undoped PrMnO$_3$ manganite has been investigated in details at various temperatures between 300 and 4 K. Its low energy spectrum exhibits an optical gap, and is characterized by a single broad peak centered at $\sim$ 2 eV. This peak is interpreted in terms of an indirect interband transition between the split bands $e_\text{g}$ caused by a strong electron Jahn-Teller phonon coupling. The spectral weight of this transition is found to be related to the magnetic ordering, which consists of ferromagnetic planes coupled antiferromagnetically. We show that such a 2D ferromagnetism plays, via the double exchange interaction, an essential role in the electronic properties of PrMnO$_3$, which is a 3D antiferromagnetic compound. Finally, an excess of optical spectral weight is found above the N\'eel temperature, and is attributed to  ferromagnetic fluctuations. A signature of such fluctuations is equally found from electron spin resonance experiments.
\end{abstract}

\section{Introduction}

The colossal magnetoresistance observed in perovskite manganese oxides has revived the interest in these systems  exhibiting a large variety of intriguing phenomena \cite{Tokura99}. These unusual physical properties arise from the interplay between interactions resulting in a rich electronic phase diagram which exhibits different charge, spin and orbital orderings \cite{Dagotto03}. In the perovskite manganites, the  Mn ions are located at the center of MnO$_6$ octahedrons and their $3d$ levels are splitted in a triplet $t_\text{2g}$ and a doublet $e_\text{g}$ in the cubic crystal field. The $t_\text{2g}$ and $e_\text{g}$ electrons are into a high spin state because of a strong Hund coupling $J_{\text{H}}$. In the undoped compounds LnMnO$_3$, where Ln is a lanthanide ion, the Mn ions are trivalent with $(t_\text{2g})^{3}(e_\text{g})^{1}$ configuration. The Jahn-Teller (JT) effect, which is very important because of the strong electron phonon coupling, removes the orbital degeneracy of the doublet. Cooperative JT effects occur below $T_{\text{JT}}$ ($\sim 900\: \text{K}$ for PrMnO$_3$ \cite{MartinCarron01}) yielding both an orbital ordering \cite{Hotta99,Murakami98} and the opening of a gap. In addition to the orbital ordering, A-type antiferromagnetic (AF) ordering, corresponding to $ab$-planes that are ferromagnetic (FM) with antiferromagnetic coupling between them, is stabilized below $T_{\text{N}}$ ($\sim100\: \text{K}$ for PrMnO$_3$ \cite{Mukhin01,Hemberger04}). Such an orbitally ordered A-type AF state has been generally explained on the basis of Goodenough-Kanamori-Anderson rules \cite{Khomskii97}, which are based on the superexchange interaction.

LaMnO$_3$ has been widely studied in recent years. In particular, it was found \cite{Quijada01,Kim04} that the low energy spectrum of the optical conductivity displays a peak at $2\: \text{eV}$ characterized by an increase of its spectral weight at low-$T$. Such a $T$-dependence was \textit{qualitatively} explained in terms of double exchange (DE) effects on optical transitions \cite{Quijada01,Kim04} in agreement with theoretical predictions \cite{Ahn_Millis00}. In this Letter, we investigate the $T$-dependence of the optical conductivity of PrMnO$_3$, which to the best of our knowledge has not been previously reported. We find a behaviour clearly similar to that reported for LaMnO$_3$ \cite{Quijada01,Kim04}. The $T$-dependence of the peak around $2\: \text{eV}$ is a robust feature of this type of coumpound, and results from the FM spin ordering in the $ab$-planes. Finally, we find \textit{quantitative} agreement with the Furukawa's double exchange model \cite{Furukawa99} extended to two $e_\text{g}$ bands splitted by cooperative JT effects.

\section{Experimental details}

The samples used in this study are polycrystallines of the stoichiometric compounds PrMnO$_3$ and CaMnO$_3$. They were synthesized from an organic gel-assisted citrate process described in \cite{Hassini05}. Reflectivity spectra were measured by using a Fourier-transform infrared (Bruker IFS 66v/S) spectrometer over the energy range $6\: \text{meV}$ to $2.2\: \text{eV}$. In the $T$ range $4\: \text{K}$ to $300\: \text{K}$, the measurements were carried out by using a He-cryostat with optical windows. The samples were polished up to optical grade ($0.25\: \mu \text{m}$). After the initial measurements, the samples were coated with a silver film and measured again. These additional data were used as reference mirrors with the purpose of taking into account of light scattering on the surface of samples. The complex dielectric function and the optical conductivity were obtained by applying Kramers - Kr$\mathrm{\ddot{o}}$nig  transformation (KK). To perform KK, the reflectance data were extrapolated at high energy using a Lorentz oscillator around $3\: \text{eV}$ \cite{TaPhuoc03}. The Electron Spin Resonance (ESR) measurements were performed on a Bruker (EMX 6/1) spectrometer operating in the X-band ($9.5\: \text{GHz}$). 

\section{Origin of the 2 eV peak}

\begin{figure}
\centerline{\epsfxsize=9 cm $$\epsfbox{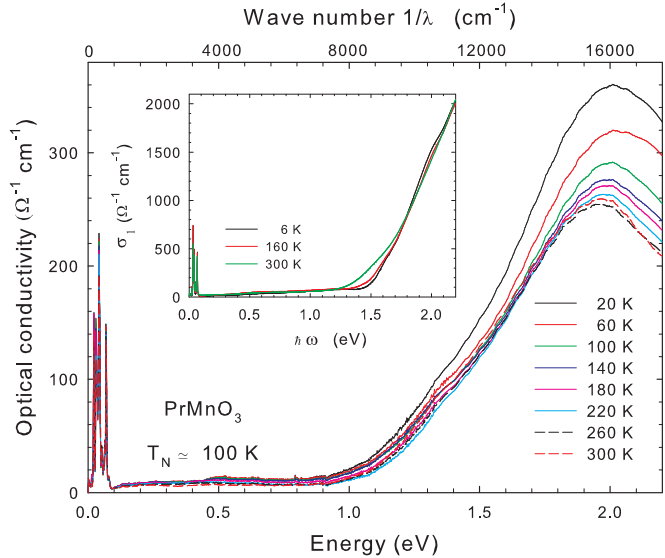}$$}
\caption{Optical conductivity of PrMnO$_3$ (main part) and CaMnO$_3$ (insert) at different temperatures.}
\label{conduc:fig}
\end{figure}

Figure \ref{conduc:fig} shows the real part of the optical conductivity, $\sigma_\text{1}(\omega)$, of PrMnO$_3$ at different temperatures. Below $0.1\: \text{eV}$, we observe severals peaks that correspond to the phononic contribution. An optical gap goes up to the onset ($\sim1\: \text{eV}$) of a peak at $\sim2\: \text{eV}$, which is characterized by three main features: $(i)$ a $T$-dependence, $(ii)$  a clear correlation with the magnetic ordering (its height grows  between $300\:\text{K}$ and $T_{\text{N}}\approx100\: \text{K}$ as much as between $T_{\text{N}}$ and $20\: \text{K}$), and $(iii)$ a large width ($\sim1\: \text{eV}$).
Several optical excitations may be considered to understand the origin of this peak : a crystal field (CF) transition ($t_\text{2g}\longrightarrow e_\text{g}$), a charge transfert transition (CT) ($O_\text{2p}\longrightarrow e_\text{g}$) or a JT transition ($e_{\text{g}_1}\longrightarrow e_{\text{g}_2}$).

A CF transition costs the energy $\Delta_\text{CF}+ E_\text{JT}$ when exciting an electron whose spin is parallel to the spin of the site or $\Delta_\text{CF}+2J_\text{H}$ when the spins are antiparallel, where $\Delta_\text{CF}$ and $E_\text{JT}$ are respectively the energy difference between $t_\text{2g}$ and $e_{\text{g}_1}$ levels, and between $e_{\text{g}_1}$ and $e_{\text{g}_2}$ levels. According to ab initio calculations on LaMnO$_{3}$ \cite{Elfimov99,Munos04,Pickett96,Ravindran02,Satpathy96}, one expects $\Delta_\text{CF}\approx1\: \text{eV}$, $E_\text{JT}\geq1\: \text{eV}$ and $2J_\text{H}\approx1-3\: \text{eV}$. It follows that such a CF transition is over our experimental window.
On the other hand, CT for an electron from $O_\text{2p}$ to $e_{\text{g}_2}$ or from $O_\text{2p}$ to $e_{\text{g}_1}$ respectively costs $\Delta_\text{CT}$ and $\Delta_\text{CT}-E_\text{JT}+2J_\text{H}$. The second transition costs more energy than the first one, because $2J_\text{H}>E_\text{JT}$. Ab initio calculations \cite{Anisimov97} lead to $\Delta_\text{CT}>3\: \text{eV}$, so we believe that such a transition is not visible in our experimental window. Moreover, CT transition does not depend on the magnetic order, and hence, it should not vary with $T$ (especially at $T_\text{N}$).
An estimation of $E_\text{JT}$ obtained from ab initio calculations \cite{Anisimov97,Elfimov99,Munos04,Pickett96,Satpathy96} is consistent with a $2\: \text{eV}$ peak in $\sigma_{1}(\omega)$. The effect of the magnetization on the transition depends on the approach (site or bands). It is unaffected within a site approach \cite{Allen99}, whereas, in a band picture, Ahn and Millis \cite{Ahn_Millis00} have shown that the part of the spectrum of $\sigma_{1}(\omega)$ related to the JT excitation increases at the transition from the paramagnetic state to the A-type AF state, due to the FM spin ordering in the $ab$-planes. In LaMnO$_3$, Quijada et al. \cite{Quijada01} and Kim et al. \cite{Kim04} have observed such an increase of the $2\: \text{eV}$ feature, below $T_\text{N}$. 

In contrary to PrMnO$_3$, the Mn ions in CaMnO$_3$ are tetravalent with $(t_\text{2g})^{3}$ configuration. No orbital degeneracy exists, and hence, JT effect does not appear. Therefore, the lowest optical interband transition should be either CT-type or CF-type. In both cases no $T$-dependence is expected \cite{Jung97,Loshkareva04,Pickett96}. This agrees with experimental results found for CaMnO$_3$, as shown in the insert of fig.\ref{conduc:fig}.

Moreover, it has been shown from ab initio calculations on LaMnO$_3$ \cite{Pickett96} that the gap between the JT splitted bands is indirect. We will now show that the $2\: \text{eV}$ feature observed in the optical conductivity spectrum of PrMnO$_3$ exhibits the typical optical properties of an indirect gap (phonon assisted transition). This last fact reinforces the hypothesis of an interband JT transition, which will be assumed in the following.

\section{Indirect gap}

\begin{figure}
\centerline{\epsfxsize=14.4 cm $$\epsfbox{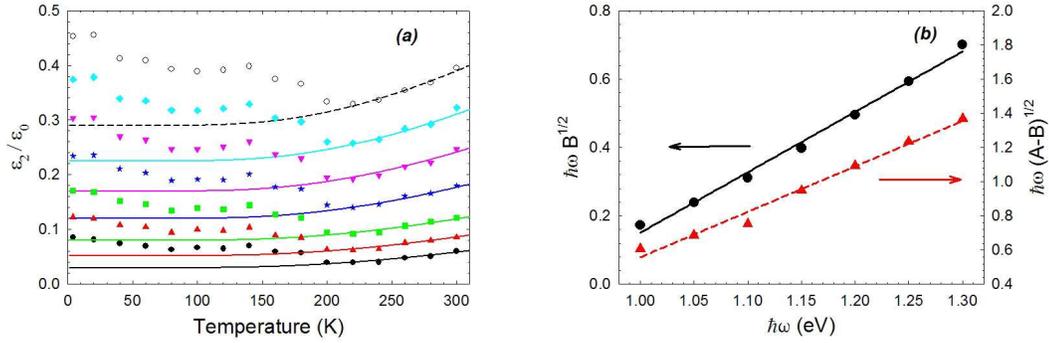}$$}
\caption{\textbf{a)} $\epsilon_2$ vs temperature for different energy values from $1\: \text{eV}$ (bottom) to $1.30\: \text{eV}$ (top) by step of $0.05\: \text{eV}$; the lines are a fit of eq.(\ref{gap_indirect:eq}) to the data. \textbf{b)} Plots of $\hbar\omega (A-B)^\frac{1}{2}$ and $\hbar\omega B^\frac{1}{2}$ vs energy; the lines are linear regressions to the data; $A$ and $B$ are defined in the text.}
\label{gap:fig}
\end{figure}

In order to follow the gap evolution, we study the $T$-dependence of the imaginary part  of the dielectric function, $\epsilon_2$, at fixed energy in the vicinity of the gap [\textit{cf.} fig.\ref{gap:fig}a]. It should be noted that the bidimensional (2D) ferromagnetism may mask the phenomena of weak strength at low-$T$ and above $T_{\text{N}}$ where magnetic fluctuations exist, and hence, an indirect gap should be more easy observable at sufficient high-$T$ if some phonon assisted optical excitations do exist.

If one considers that a single phonon mode participates in the combined electron-radiation and electron-phonon interactions, $\epsilon_2 (\omega,T)$ is given by (see for example ref. \cite{Harbeke72}):
 
\begin{eqnarray}
\epsilon_2=\frac{1}{\omega^2}\left[\left|\chi_\text{abs}\right|^2 N_\text{BE}(T)\:(\hbar \omega - E_\text{g} + \hbar \omega_\text{ph})^2 + \left|\chi_\text{emis}\right|^2(N_\text{BE}(T)+1)\:(\hbar \omega - E_\text{g} - \hbar\omega_\text{ph})^2\right]
\label{gap_indirect:eq}
\end{eqnarray}

In eq.(\ref{gap_indirect:eq}), the first and the second terms respectively characterize the light absorption process involving the absorption and the emission of phonons of energy $\hbar\omega_\text{ph}$ whose average number is given by the Bose-Einstein distribution function $N_\text{BE}(T)=1/(\exp[\hbar\omega_\text{ph}/k_\text{B}T]-1)$, where the chemical potential is zero. We see that the threshold for indirect transition is at $\hbar\omega =E_\text{g}-\hbar\omega_\text{ph}$ where $E_\text{g}$ is the gap energy. Note that the $T$-dependence of the two components arises from $N_\text{BE}(T)$ since $\chi_\text{abs}$ and $\chi_\text{emis}$, which respectively represent the matrix elements of the electron-radiation-phonon interaction for the phonon absorption process and for the phonon emission process, are independent of $T$.

In fig.\ref{gap:fig}a, we plot $\epsilon_2$ as a function of $T$ at $\omega$ kept constant. There seems to be no doubt that the $T$-dependence of $\epsilon_2$ at high-$T$ is governed by $N_\text{BE}(T)$, as may be seen in fig.\ref{gap:fig}a where the lines display a fit of $A\:N_\text{BE}(T) + B$ to the data. The only free parameters are $A$, $B$ and $\omega_\text{ph}$. According to eq.(\ref{gap_indirect:eq}), $A$ and $B$ are respectively related to $\vert\chi_\text{emis}\vert^2/\omega^2\:(\hbar\omega - E_\text{g} - \hbar\omega_\text{ph})^2 + \vert\chi_\text{abs}\vert^2/\omega^2\:(\hbar\omega - E_\text{g} + \hbar\omega_\text{ph})^2$, and to $\vert\chi_\text{emis}\vert^2/\omega^2\:(\hbar\omega - E_\text{g} - \hbar \omega_\text{ph})^2$. The best fit is obtained for $\hbar\omega_\text{ph} = 0.07\: \text{eV}$, which is a realistic phonon energy corresponding to stretching vibration modes of the MnO$_{6}$ octahedron. The $\omega$-dependence of the two other fitting parameters are shown in fig.\ref{gap:fig}b, where $\hbar\omega\:(A-B)^\frac{1}{2}$ and $\hbar\omega\:B^\frac{1}{2}$ are plotted as a function of $\hbar\omega$. In both cases, we obtain a linear law which means that $\chi_\text{abs}$ and $\chi_\text{emis}$ display only a small $\omega$-dependence in the vicinity of the band gap. Finally, we obtain $E_\text{g}=0.83\pm0.01\: \text{eV}$.
  
\section{Spectral weight analysis}

\begin{figure}
\centerline{\epsfxsize=9 cm $$\epsfbox{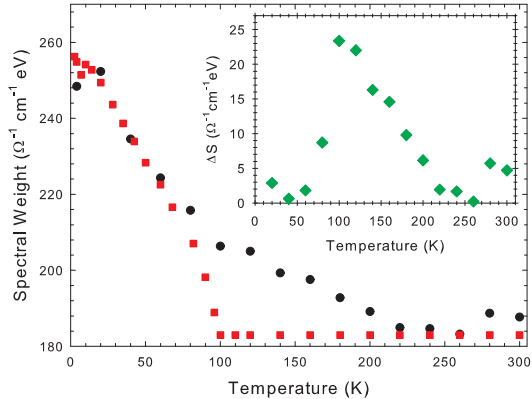}$$}
\caption{Temperature dependence of the spectral weight of the electronic contribution of $\sigma_{1}(\omega)$ of PrMnO$_3$; the solid squares are a fit of eq.(\ref{total_S:eq}) to the data (solid circles). Insert: $\Delta S$ vs temperature.}
\label{SW_prmno3}
\end{figure}

We shall now pass on a quantitative investigation of the $2\: \text{eV}$ feature at low-$T$. Hereafter, we define the integrated spectral weight of the electronic contribution of $\sigma_{1}(\omega)$, i.e. $S\equiv\int^{2.2}_{0.1}\sigma_{1}(\omega)\: \text{d}\omega$ where we neglect the area below $0.1\: \text{eV}$ because it originates essentially from phononic contributions. Solide circles in fig.\ref{SW_prmno3} display the $T$-dependence of $S$. We see that $S$ increases as $T$ decreases, exhibiting a variation rate larger at $T<T_\text{N}\approx 100\: \text{K}$ than at $T>T_\text{N}$. Such $T$-dependent changes in $S$ have already been observed in LaMnO$_3$, and have qualitatively been explained in terms of DE picture \cite{Quijada01}. According to this view, we succeed in interpreting quantitatively the results presented here by making an extension of recent theoretical findings obtained for DE systems \cite{Furukawa99}.

Furukawa has calculated the density of state (DOS) at finite $T$ of the (one-orbital) DE model using non-perturbative approaches \cite{Furukawa99}. He found that the DOS is characterized by a two-peak structure (reflecting the two subband structure $e_\text{g}^\text{(u)}$ and $e_\text{g}^\text{(l)}$ at $\sim\pm J_\text{H}$ due to strong FM Hund coupling $J_\text{H}\gg W$, where $W$ is the one-electron bandwidth of the $e_\text{g}$ band) whose $T$-dependence is determined by the magnetization through $(1+M)/2$ [$(1- M)/2$] for the up [down] spin electrons in the lower spin subband $e_{\text{g}_{\uparrow}}^\text{(l)}$ [$e_{\text{g}_{\downarrow}}^\text{(l)}$] and for the down [up] spin electrons in the upper spin subband $e_{\text{g}_{\downarrow}}^\text{(u)}$ [$e_{\text{g}_{\uparrow}}^\text{(u)}$]. So, the lower subband $e_\text{g}^\text{(l)}$ and the upper subband $e_\text{g}^\text{(u)}$ are completely symmetric in the paramagnetic state ($M=0$), while $e_{\text{g}_{\uparrow}}^\text{(l)}$ and $e_{\text{g}_{\downarrow}}^\text{(u)}$ [$e_{\text{g}_{\downarrow}}^\text{(l)}$ and $e_{\text{g}_{\uparrow}}^\text{(u)}$] become majority [minority] spin bands below the Curie temperature. Finally, perfect spin polarization appears at the FM ground state ($M=1$) in which $e_\text{g}^\text{(l)}$ is only occupied by electrons with up spin (parallel to $M$), whereas down spin states only exist in $e_{g}^{(u)}$. We extend these results by considering the two active $e_\text{g}$ orbitals for large electron JT phonon coupling $E_\text{JT}\gg W$. Within the framework of this approximation, one expects that the DOS is split into four parts, which reflect for $J_\text{H}>E_\text{JT}/2$ two lower subbands $e_{\text{g}_{1}}^\text{(l)}$ and $e_{\text{g}_{2}}^\text{(l)}$ at $-J_\text{H}\mp E_\text{JT}/2$ and two upper subbands $e_{\text{g}_{1}}^\text{(u)}$ and $e_{\text{g}_{2}}^\text{(u)}$ at $J_\text{H}\mp E_\text{JT}/2$. Hereafter, we confine ourselves only to $e_{\text{g}_{1}}^\text{(l)}$ and $e_{\text{g}_{2}}^\text{(l)}$. Their DOS is proportional to $(1\pm M)/2$ for electrons with up (down) spin. According to the Fermi golden rule, the optical transition rate of $e_{\text{g}_{1}}^\text{(l)}\rightarrow e_{\text{g}_{2}}^\text{(l)}$ should be proportional to the product of the DOS for the initial state by the DOS for the final state. So, taking into consideration the only allowed interband excitations $e_{\text{g}_{1 \uparrow}}^\text{(l)}\rightarrow e_{\text{g}_{2 \uparrow}}^\text{(l)}$ and $e_{\text{g}_{1 \downarrow}}^\text{(l)}\rightarrow e_{\text{g}_{2 \downarrow}}^\text{(l)}$ in accordance with standard selection rule forbidding optical transitions between states with opposite spins, we obtain that $S$ should be proportional to $(1+M^{2})/2$. 

In fact, such an evaluation of $S$ holds solely for $S_{ab}$ originating from intraplanar electronic excitations, since the magnetic stucture of PrMnO$_3$ consists of FM $ab$-planes coupled antiferromagnetically between them. The $c$-axis contributions to $S$ would have to come from optical transitions from initial states lying within majority (minority) spin subband $e_{\text{g}_{1 \sigma}}^\text{(l)}$ to final states lying within minority (majority) spin subband $e_{\text{g}_{2 \sigma}}^\text{(l)}$. It follows that such interplanar electronic processes should result in a spectral weight $S_{c}$ proportional to $(1-M^{2})/2$, where $M$ is henceforth the reduced magnetization within the $ab$-planes. We see that $S_{c}$ decreases as $M$ is increased below $T_\text{N}$, and vanishes at the ground AF state. This is a direct consequence of the spin polarization effect which tends to cancel out the probability of interplanar electronic excitations. We thus obtain that the total spectral weight $S$ should be given by:
\begin{eqnarray}
S=\;\alpha_{ab}\:S_{\infty}\:(1+M^{2})\;+\;\alpha_{c}\:S_{\infty}\:(1-M^{2})
\label{total_S:eq}
\end{eqnarray}
where $S_{\infty}$ is the spectral weight at high-$T$ ($M=0$), and $\alpha_{ab}$ and $\alpha_{c}$ are weighted factors such that $\alpha_{ab}+\alpha_{c}=1$. For a system of cubic symetry, one expects $\alpha_{ab}=2\alpha_{c}=2/3$. 

Using the measurements of $M(T)$ published in \cite{Mukhin01}, we have fitted the eq.(\ref{total_S:eq}) to the data. Note that there exits only two independent fitting parameters, as may be seen by rewritting the above formula in the form $S_{\infty}\:(1+C\:M^{2})$ where $C$ is related to $\alpha_{ab}-\alpha_{c}$.  In fig.\ref{SW_prmno3}, the solid squares display the best fit, whence we obtain $\alpha_{ab}\approx0.7$ and $\alpha_{c}\approx0.3$. These values are consistent with the expected orthorhombic structure ($Pnma$ space group) which originates from the ideal cubic structure due to cooperative JT effects occuring below $T_\text{JT}\sim900\: \text{K}$ \cite{MartinCarron01}. 

A difference may be noted between the predictions of the model and the observed spectral weight $S_\text{exp}$ for $T>T_\text{N}$ up to $\sim2\:T_\text{N}$. Such a discrepancy between $S_\text{exp}$ and the spectral weight calculated from eq.(\ref{total_S:eq}) [\textit{cf.} insert of fig.\ref{SW_prmno3}] can be understood in terms of an excess of spectral weight, $\Delta S=S_\text{exp}-S$, due to FM fluctuations, as it will be seen below.  
\section{Ferromagnetic fluctuations} 

\begin{figure}
\centerline{\epsfxsize=14.4 cm $$\epsfbox{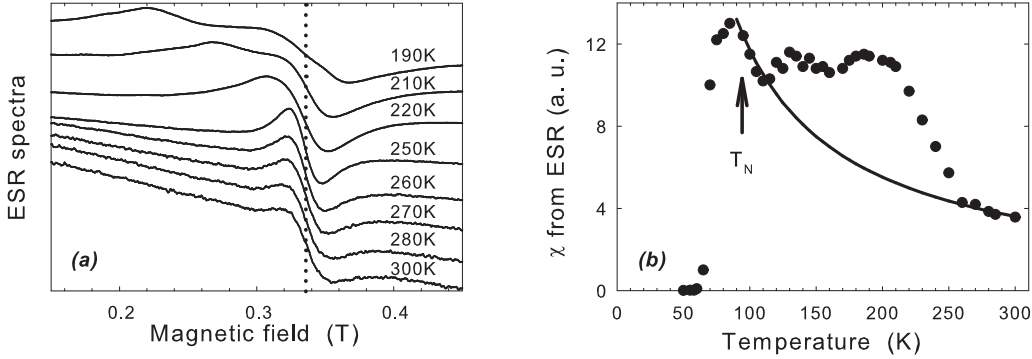}$$}
\caption{\textbf{a)} Evolution with temperature of ESR spectra of PrMnO$_3$; the curves are rescaled to have the same peak-to-peak amplitude, and are shifted with respect to one another along the ordinate axis; the vertical dotted line shows the resonance field of the central peak.  \textbf{b)} The spin susceptibility as deduced from the integral of ESR absortion spectra, the line is a Curie-Weiss law fit to the data.}
\label{esr:fig}
\end{figure}

We have measured ESR absorption spectra at $T>T_{\text{N}}$. An important difference may be noted between the spectra recorded above $250\: \text{K}$ and those obtained below $250\: \text{K}$. In the $T$ range $250\: \text{K}$ to $300\: \text{K}$, we observe a single absorption peak, centered around a resonance field $H_\text{r}\approx0.335\: \text{T}$, implying a Land\'{e} factor $g\approx2.02$ characteristic of the paramagnetic (insulating) phase, whereas, below $250\: \text{K}$, a two-feature-like structure appears on either side of the central peak, which remains centered around $H_\text{r}$. It may be seen from fig.\ref{esr:fig}a, where the ordinate axis represents the derivative of absorption with respect to magnetic field, that the low-field feature moves towards lower field values as $T$ decreases, while the higher-field feature shifts towards higher field values. Such a behaviour of ESR spectra may be explained by the appearance of ferromagnetically ordered regions in the sample below $250\: \text{K}$ \cite{Reynaud00}. A possible cause of this weak ferromagnetism is the existence of 2D FM fluctuations due to the orbital ordering. Such a scenario is in agreement with observations reported in LaMnO$_{3}$ \cite{Tovar99}. Furthermore, the integral of the absorption over the magnetic field range studied, which is proportional to the spin susceptibility $\chi$, follows a Curie-Weiss law $\chi_\text{CW}=C/(T-\theta)$ with $\theta=11\: \text{K}$ in the $T$ range $250\: \text{K}$ to $300\: \text{K}$, while an important enhancement of $\chi$ occurs below $250\:K$ due to FM fluctuations, as shown in fig.\ref{esr:fig}b. From these results, we may notice the correlation between the evolution with $T$ of the difference $\chi-\chi_\text{CW}$ and the one of the excess of spectral weight, $\Delta S$, shown in the insert of fig.\ref{SW_prmno3}. This indicates that the 2D FM fluctuations are good candidates to understand the origin of $\Delta S$

\section{Conclusion}
We have investigated the evolution with $T$ of the optical conductivity spectrum for a polycrystalline of PrMnO$_{3}$. We have observed a single optical excitation around $\sim2\: \text{eV}$, that we interpret by considering an indirect interband transition between the two lower subbands arising from the splitting of $e_\text{g}$ states due to both strong FM exchange coupling and large electron JT phonon coupling. The $T$-dependence of the spectral weight of the transition is quantitatively explained by eq.(\ref{total_S:eq}), that we have established from an extension of recent theoretical results obtained for DE ferromagnets \cite{Furukawa99}. These results leads us to propose that the DE interaction plays an important role not only in the metallic ferromagnetic state of hole-doped manganites but also in PrMnO$_{3}$, which is an insulator A-type antiferromagnet at low-$T$.

\acknowledgements
We acknowledge the financial support provided through the CAPES-COFECUB program under contract No. 500/05.

\end{document}